\documentclass[a4paper,12pt]{article}  % other types include report, book, letter, slides

\usepackage{amsmath,amssymb}
\usepackage{setspace}
\usepackage[left=1.5cm,top=2cm,bottom=2.5cm,right=1.5cm,nohead]{geometry}
\usepackage{graphicx}
\usepackage{alltt}
\usepackage{listings}
\usepackage{dcolumn}
\usepackage{bm}
\usepackage{hyperref}
\usepackage{amsmath,amssymb}

\newcommand{\bt}[1]{{\mathbf #1}}

\begin{document}
\title{\Large \textbf{Counting of discrete Rossby/drift wave resonant triads (again)}}
\author{Miguel D. Bustamante${}^{1,}$\footnote{E-mail: \texttt{miguel.bustamante@ucd.ie}}
, Umar Hayat${}^2$, Peter Lynch${}^1$, Brenda Quinn${}^1$\\
{\small ${}^1$ School of Mathematical Sciences,
                University College Dublin,  Belfield, Dublin 4, Ireland}\\
{\small ${}^2$ Department of Mathematics, Quaid-i-Azam University,  Islamabad 45320, Pakistan}}

\maketitle

\bibliographystyle{plain}  % want neuron

\abstract{The purpose of our earlier note \cite{BHLQ01} was to remove the
confusion over counting of resonant wave triads for Rossby and drift waves
in the context of the Charney-Hasegawa-Mima equation. A comment
\cite{Kartashov2013B} on that note has further confused the situation.
The present note aims to remove this obfuscation.
}

\section{Counting of Triads}

In our earlier note \cite{BHLQ01}, we pointed out a significant error
of over-counting of triads in \cite{[Kartashov and Kartashova 2013]}.
For a \emph{real} field $\psi(x,y,t)$ the modes $\hat{\psi}_{\bt k}$ and
$\hat{\psi}_{-{\bt k}}$ must occur with amplitudes which are complex
conjugates. They are not independent.
Thus, the claim of \cite{[Kartashov and Kartashova 2013]} that they are listing six separate
triads in their equation (10) is wrong. In fact, all six triads are equivalent.
A comment \cite{Kartashov2013B} on our note has once more confused the situation.

\section{Real Fields, not Real Coefficients}

The comment of \cite{Kartashov2013B} claims that ``\cite{BHLQ01} states that \cite{[Kartashov and Kartashova 2013]}
counted triads with \emph{real} amplitudes.'' This is wrong. In fact,
\cite{BHLQ01} states that the underlying field $\psi$ is real and that,
consequently, the amplitudes occur in complex conjugate pairs.

Again, in their Conclusions,
\cite{Kartashov2013B} write that ``\cite{BHLQ01} did not notice that the dynamical
system (4) in \cite{[Kartashov and Kartashova 2013]} is written in complex variables.'' This is
nonsense. It is abundantly clear that the variables in (4) are complex: there
is no misunderstanding on this point. It is the physical field that is real.

%%%%%%%%%%%%%%%%%%%%%%%%%%%%%%%%%%%%%%%%%%%%%

\section{Deduction by \cite{[Kartashov and Kartashova 2013]} from a Theorem in \cite{[Yamada and Yoneda 2013]}}

\cite{[Kartashov and Kartashova 2013]} use a result of \cite{[Yamada and Yoneda 2013]}, which is valid in the
asymptotic limit of infinitely large $\beta$, to deduce results for finite
$\beta$. An asymptotic result like this should be valid as
$\beta\longrightarrow\infty$, but is invalid for small or moderate $\beta$, as in
\cite{Bustamante2013}. The deductions of \cite{[Kartashov and Kartashova 2013]} are unsustainable.

\section{Conclusions}

In \cite{Kartashov2013B} there is confusion between what is real and what is complex.
This has once again confused the picture that we were aiming to clarify in
\cite{BHLQ01}.  The present note should remove this obfuscation.

The Appendix contains important facts that provide evidence of the completeness of the classification of discrete resonant triads for Rossby/drift waves in periodic domains presented by the published paper \cite{Bustamante2013}, regardless of the unfounded comments by Kartashov and Kartashova in the preprints \cite{[Kartashov and Kartashova 2013],Kartashov2013B}.

\section{Acknowledgments}
We thank Sergey V. Nazarenko for useful discussions.

%%%%%%%%%%%%%%%%%%%%%%%%%%%%%%%%%%%%%%%%%%%%%%%%%

%\bibliography{bibliography}

\section*{Appendix: Three Important Facts}
\noindent \textbf{Fact 1.} The classification of discrete resonant triads for Rossby/drift waves in periodic domains presented by the published paper \cite{Bustamante2013} is in fact \emph{complete}. The reason for the completeness is that the classification is based on an explicit bijective mapping from the set of non-zonal irreducible triads to the set of representations of integers as sums of the form $r^2 + s^2$ and $3\, m^2 + n^2.$ The explicit representations are due to well-known results by Pierre de Fermat and applied in \cite{Bustamante2013} for the first time to the problem of finding exact resonances for the Charney-Hasegawa-Mima equation in periodic domains.\\
\noindent \textbf{Fact 2.} There is a repeated critique in \cite{[Kartashov and Kartashova 2013],{Kartashov2013B}} about the minimum level of detuning $\delta_{\min}$ of the set of quasi-resonant triads found by \cite{Bustamante2013} in the box of size $L=100$ ($\delta_{\min} \approx 10^{-5}$ as compared to $\delta_{\min} \approx 10^{-12}$ of the brute-force search). This is not an issue since \cite{Bustamante2013} explicitly state (p.~2414) that they sample only a fraction of the total available triads ($40 \,434$ out of $\approx L^4$ triads) in order to study the clusters' connectivity properties and percolation transition, all as functions of the allowed detuning. The success of \cite{Bustamante2013} is that the observed features (connectivity and percolation) \emph{for such small sample} have the same properties as the 
corresponding features \emph{for the whole set of triads}, computed directly by brute force in a thorough study at higher resolution ($L \geq 256$) published by Bustamante and coworkers in \cite{HCB13}.\\
\noindent \textbf{Fact 3.} The quasi-resonances found by \cite{Bustamante2013} are close to resonant manifolds, \emph{not to exact discrete resonances}. Preprint \cite{Kartashov2013B} introduced confusion again into the subject by citing extracts from \cite{Bustamante2013} and interpreting them literally. Let us clarify the matter: an arbitrary point $(\mathbf{k}_1, \mathbf{k}_2)$ on a resonant manifold is generically an ``exact resonance'' in the sense that it still satisfies $\mathbf{k}_1 + \mathbf{k}_2 = \mathbf{k}_3$ and $\omega(\mathbf{k}_1) + \omega(\mathbf{k}_2) = \omega(\mathbf{k}_3),$ but with non-integer wavevectors. This is not to be confused with an \emph{exact discrete resonance} (the matter of our research, corresponding to a set of integer wavevectors). In order to make the wavevectors physically sensible we move the point to a nearby integer point. In so doing, the equations $\mathbf{k}'_1 + \mathbf{k}'_2 = \mathbf{k}'_3$ are maintained for the new integer wavevectors but now $\omega(\mathbf{k}'_1) + \omega(\mathbf{k}'_2) \neq \omega(\mathbf{k}'_3),$ providing thus a quasi-resonant triad lying close to the resonant manifold, \emph{not close to an exact discrete resonance}. For example, one of the exact discrete resonant triads we use for building a set of quasi-resonances in the box of size $L=100$ is the triad $\mathbf{k}_1 = \{11\, 171\, 680, 463\, 515 \,988\}, \mathbf{k}_2 =  \{990 \,044 \,945, -305 \,135 \,237\},$ $\mathbf{k}_3 = \mathbf{k}_1 + \mathbf{k}_2 = \{1 \,001 \,216 \,625, 158 \,380 \,751\}.$ This triad is irreducible (the set of six components is relatively prime), satisfies $\omega(\mathbf{k}_1) + \omega(\mathbf{k}_2) = \omega(\mathbf{k}_3)$ and is $10^7$ times greater than the box. Our construction of quasi-resonant triads out of this ``generating triad'' is as follows: (i) Re-scale this triad by a common real factor $\alpha$ so that the re-scaled non-integer triad $\alpha \mathbf{k}_1, \alpha \mathbf{k}_2, \alpha \mathbf{k}_3$
fits into the $L=100$ box. (ii) Replace this non-integer triad with a nearby integer triad that satisfies $\mathbf{k}'_1 + \mathbf{k}'_2 = \mathbf{k}'_3.$ This will automatically satisfy $\omega(\mathbf{k}'_1) + \omega(\mathbf{k}'_2) \neq \omega(\mathbf{k}'_3),$ so the new triad is quasi-resonant. (iii) Repeat part (ii) with several nearby integer triads. (iv) Repeat part (i) using all allowed values of $\alpha$. This produces about $1 000$ quasi-resonant triads in the box $L=100$ out of the generating triad $\mathbf{k}_1, \mathbf{k}_2, \mathbf{k}_3.$ Finally, start again with another big irreducible generating triad. The whole process' computational time scales like $L^2$ and is thus very efficient as compared to a brute-force search of \emph{all quasi-resonant triads}, where computational time scales like $L^4$ (and even slower due to storage and sampling issues) in the box of size $L$.


\begin{thebibliography}{1}

\bibitem[BH13]{Bustamante2013}
M.~D.~Bustamante and U.~Hayat.
\newblock Complete classification of discrete resonant Rossby/drift wave triads
  on periodic domains.
\newblock {\em Commun. Nonlinear Sci. Numer. Simulat.}, 18:2402 -- 2419, 2013.

\bibitem[BHLQ13]{BHLQ01}
M.~D.~Bustamante, U.~Hayat, P.~Lynch and B.~Quinn.
\newblock Counting of discrete Rossby/drift wave resonant triads.
\newblock {\tt arXiv:1309.0405 [physics.flu-dyn]}

\bibitem[HCB13]{HCB13}
J.~Harris, C.~Connaughton and M.~D.~Bustamante.
\newblock Percolation transition in the kinematics of nonlinear resonance broadening in Charney-Hasegawa-Mima model of Rossby wave turbulence.
\newblock  {\em New
Journal of Physics}, 15, 083011, 2013.

\bibitem[KK13]{[Kartashov and Kartashova 2013]}
A.~Kartashov and E.~Kartashova.
\newblock Discrete exact and quasi-resonances of Rossby/drift waves on $\beta$-plane with periodic boundary conditions.
\newblock {\tt arXiv:1307.8272v1 [physics.flu-dyn]}

\bibitem[KK13a]{Kartashov2013B}
A.~Kartashov and E.~Kartashova.
\newblock Comment to the note ``Counting of discrete Rossby/drift wave resonant triads''.
\newblock {\tt arXiv:1309.0992v1 [physics.flu-dyn]}

\bibitem[YY13]{[Yamada and Yoneda 2013]}
M.~Yamada and T.~Yoneda.
\newblock Resonant interaction of Rossby waves in two-dimensional flow on a
  plane.
\newblock {\em Physica D: Nonlinear Phenomena}, 245:1 -- 7, 2013.




\end{thebibliography}
\end{document}